\documentclass[11pt]{article}
\def\be{\begin{equation}}
\def\ee{\end{equation}}
\def\ba{\begin{eqnarray}}
\def\ea{\end{eqnarray}}
\def\la{\langle}
\def\ra{\rangle}
\def\a{\alpha}

\def\h{\hskip 1cm}

\usepackage{epsfig}
\begin{document}
\begin{titlepage}
\vspace{4cm}
\begin{center}{\Large \bf Transition behavior in the capacity of correlated-noisy channels in arbitrary dimensions}\\
\vspace{1cm}V. Karimipour \footnote{Corresponding author,
email:vahid@sharif.edu}\\ L.
Memarzadeh\footnote{email:laleh@mehr.sharif.edu}, \\
\hspace{0.5cm}
\vspace{1cm} Department of Physics, Sharif University of Technology,\\
P.O. Box 11365-9161,\\ Tehran, Iran
\end{center}
\vskip 3cm
\begin{abstract}
We construct a class of quantum channels in arbitrary dimensions for
which entanglement improves the performance of the channel. The
channels have correlated noise and when the level of correlation
passes a critical value we see a sharp transition in the optimal
input states (states which minimize the output entropy) from
separable to maximally entangled states. We show that for a subclass
of channels with some extra conditions, including the examples which
we consider, the states which minimize the output entropy are the
ones which maximize the mutual information.

\end{abstract}
\end{titlepage}

\section{Introduction}
One of the basic problems in quantum information theory \cite{Nil,
M,Holevo,Schumacher,Shor} concerns the issue of additivity of
classical capacity of quantum channels. A basic question is whether
the use of entangled states as optimal states for encoding classical
information can increase the capacity of a channel or not. A proper
calculation of the so called Schumacher-Westmoreland-Holevo capacity
\cite{Holevo},\cite{Schumacher} requires the optimization of mutual
information between the input and output of the channel when we
encode the information into arbitrary long strings of quantum states
(more precisely states in the tensor product of the Hilbert space of
one state) and carrying out the limiting procedure
$C:=\lim_{n\rightarrow\infty} C_n$, where
\begin{equation}\label{1}
    C_n:=\frac{1}{n}Sup_{\varepsilon}I_n(\varepsilon)
\end{equation}
is the capacity of the channel when we send $n-$ strings of
quantum states into the channel. Here
$\varepsilon:=\{p_i,\rho_i\}$ is the ensemble of input states,
\begin{equation}\label{mutual}
    I_n(\varepsilon):=
    S(\mathcal{E}(\sum_ip_i\rho_i))-\sum_{i}p_iS(\mathcal{E}(\rho_i))
\end{equation}
is the mutual information between the input and the output, when the
channel maps each input state $\rho_i$ into the output state
$\mathcal{E}(\rho_i)$, and $S(\rho)\equiv -tr(\rho \log \rho)$ is
the von Neumann entropy of a state $\rho$. We should stress that
(\ref{1}) may not be the proper definition for capacity in a memory
channel as formulated in its most general setting in \cite{Wer},
however such a definition seems to be still valid for generic memory
channels \cite{Wer}.\\

Calculation of $C$ is extremely difficult if not impossible. A much
simpler problem is to calculate $C_2$, and to see if entangled
states can enhance this kind of limited capacity or not. This
problem has been tackled by many authors
\cite{M1,K1,K2,M2,Matsu,Auden,M3,C} and there is now strong support
for a conjecture that if the noise of the channel is not correlated,
i.e for product channels, entangled states have no advantage over
separable states for encoding classical information. However when
the noise is correlated, several examples have been provided which
indicate that entanglement can enhance the mutual information, if
the correlation is above a certain critical value. To our knowledge
these examples are limited to qubit channels \cite{M2},\cite{M3},
and bosonic Gaussian channels
\cite{C}. \\

It is a nontrivial problem to find examples which show such a
transition. The difficulty in finding more examples resides in
the large number of parameters over which the required
optimization should be carried out. In fact one has to propose a
channel and a certain type of correlation and only after carrying
out the optimization for all values of the noise parameters and
correlation values one can see if a critical value of correlation
exists above which entangled states are advantageous over
separable states. When one goes to higher dimensional states, the
number of parameters increases and the problem becomes even more
intractable. Therefore it is desirable to have a systematic
method for constructing such channels. This is a problem which we
study in this paper.\\

We will find a general class of channels for which  there is a sharp
transition, reminiscent of phase transitions, at which the optimal
state changes abruptly from a separable state to a maximally
entangled state. This transition always happens regardless of the
value of noise parameters. Therefore it is remarkable that the
struggle for optimality is between these two extremes of
entanglement and not
between other intermediate values.\\

The structure of this paper is as follows: In section
(\ref{performance}) we first use the minimum output entropy as a
parameter characterizing the performance of the channel
\cite{Giovannetti}. Since an input state usually becomes entangled
with the environment at the output and the state of the environment
is not accessible, minimum entropy at the output means that minimum
information has leaked to the environment. \\

We then provide a general class of channels which are guaranteed to
show such transitions in their performance. In section
(\ref{capacity}) we show that for a subclass of these channels, i.e.
those for which the Kraus operators form an irreducible
representation of a group and commute modulo a phase with each
other, the minimization of output entropy is equivalent to the
maximization of the mutual information, hence we show that in the
models which we study, which are among the above subclass, it is
really the mutual information that behaves non-analytically. In
section (\ref{qubit}) we study a Pauli channel for qubits with only
bit-flip and phase flip errors and show that it shows such a
transition. Finally in section (\ref{arbitrary}) we go to arbitrary
dimensions and study a subclass of generalized Pauli channels having
a symmetry. For this subclass, still satisfying the group
representation property, we do the main part of the analysis
analytically and only at the end use numerical
calculations. Figures (1) and (2) show some of our results.\\

We should be clear that these and previous similar results on
enhancement of mutual information do not directly address the
problem of additivity of entropy, since this property deals with
sending entangled states over a product channel and not a correlated
one.

\section{Correlated channels with entanglement-enhanced
performance}\label{performance}

In arbitrary dimensions consider the following two channels, each of
them acts on a single quantum state:
\begin{eqnarray}\label{Kraus}
\Phi(\rho)&=&\sum_{\a}p_{\a}U_{\a}\rho {U_{\a}}^{\dag},\cr
\Phi^*(\rho)&=&\sum_{\a} p_{\a} U_{\a}^*\rho {U_{\a}^*}^{\dag}.
\end{eqnarray}
We take the Kraus operators \cite{K} $U_{\a}$, to be unitary
operators and $\sum_{\a} p_{\a}=1$. We now consider the channel
$\mathcal{E}$ acting on two states as follows:

\begin{equation}\label{Erho}
    \mathcal{E}(\rho):=(1-\mu)(\Phi\otimes \Phi^*)(\rho)+\mu \Phi^c(\rho),
\end{equation}
where

\begin{equation}\label{Phi}
\Phi^c(\rho)=\sum_{\a} p_{\a}(U_{\a}\otimes U_{\a}^*)
\rho(U_{\a}\otimes U_{\a}^*)^{\dag}.
\end{equation}

This type of correlation is inspired by the work of \cite{M2} who
first proposed it for the Pauli channels. The basic difference
however is that our product channel (for $\mu=0$) is $\Phi\otimes
\Phi^*$ rather than $\Phi\otimes \Phi$, and for $\mu=1$ error
operators are of the form $U_{\a}\otimes U_{\a}^*$ instead of
$U_{\a}\otimes U_{\a}$. One can not however attach the same physical
interpretation as authors of \cite{M2} did for this kind of channel,
that is one can not interpret this channel as two consecutive uses
of the channel $\Phi(\rho)$ in equation (\ref{Kraus}). However this
practical problem is of minor importance, as long as we are
interested in the non-analytical
behavior of this channel.\\

To study the performance of this channel we use minimum output
entropy as explained in the introduction. That is we search for
input states which give the minimum entropy among all output
states.\\

We know that when $\mu=0$, the channel (\ref{Erho}) is a product
channel, and for these channels, there is strong analytical and
numerical support \cite{K2} that the optimal input states of the
two channels when multiplied by each other give the best input
state of the product channel in the sense that it gives the
minimum output entropy.  On the other hand we know that when
$\mu=1$, the maximally entangled states pass through the channel
without any distortion. This is due to the easily proved identity
that for any such state, namely for any state of the form
\begin{equation}\label{}
    |\psi\ra=\frac{1}{\sqrt{d}}\sum_{i=1}^d|i,i\ra,
\end{equation}
and for any unitary operator $U$
\begin{equation}\label{}
    U\otimes U^*|\psi\ra = |\psi\ra.
\end{equation}
Thus we have $\Phi^c(|\psi\ra)=|\psi\ra$ and
$S(\Phi^c(|\psi\ra))=0$.

However this does not by itself prove that at $\mu=1$ separable
states may not be optimal, since they may have vanishing output
entropy at $\mu=1$, too. If we can prove that at $\mu=1$, no
separable state gives a vanishing output entropy, then we
conclude that somewhere in the interval $[0\ ,\ 1]$, a transition
occurs in the behavior of the channel. This by itself does not
imply that this transition should be abrupt, however our examples
clearly indicate
such an abrupt transition.\\

In order to present our next result, we need a definition. Let us
choose an arbitrary Kraus operator say $U_{\a_0}$ from the set of
$U_{\a}$'s describing the channel. For each $\a$ we define the
set of states which are invariant modulo a phase under the action
of $U_{\a}^{-1}U_{\a_0}$, that is
\begin{equation}\label{Ialpha}
I_{\a}:=\{|\psi\ra\in \mathcal{H} \ \ \ | \ \ \
U_{\alpha}^{-1}U_{\a_0}|\psi\ra=e^{ic_{\a}}|\psi\ra\}.
\end{equation}

Then the condition under which a transition occurs is simply given in the following\\

\textbf{Theorem}: If $\bigcap_{\a} I_{\a}=\O$ then there is a
transition in the optimal input states, which improve the channel
performance, from separable to maximally entangled states as we
increase the level of correlation.\\

\textbf{Proof}: We want to show that under this condition no
separable input state leads to an output state of vanishing entropy.

Let $\rho^*$ be a general separable state. By definition it can be
written as a convex combination of pure product states, namely
$\rho^*=\sum_{i}p_i \rho_i^{(1)}\otimes \rho_i^{(2)}$, where
$\rho_i^{(1)} $ and $\rho_i^{(2)}$ are pure. If
$S(\Phi^c(\rho^*))=0$, then we find from concavity of $S$ that
\begin{equation}\label{}
    S(\Phi^c(\rho^*))=S(\Phi^c(\sum_i p_i \rho_i^{(1)}\otimes
    \rho_i^{(2)}))=S(\sum_i p_i\Phi^c(\rho_i^{(1)}\otimes
    \rho_i^{(2)}))\geq \sum_i p_i S(\Phi^c(\rho_i^{(1)}\otimes
    \rho_i^{(2)})).
\end{equation}

In view of positivity of $S$ this means that for all $i$,
$S(\Phi^c(\rho_i^{(1)}\otimes \rho_i^{(2)}))=0$. Thus we should
consider only separable states of the form $\rho_1\otimes \rho_2$
where $\rho_1$ and $\rho_2$ are both pure.\\
When $\mu=1$ such a state transforms to the output state
\begin{equation}
\Phi^c(\rho_1\otimes \rho_2)=\sum_{\a} p_{\a} (U_{\a} \rho_1
{U_{\a}}^{\dag}) \otimes (U_{\a}^* \rho_2 {{U_{\a}}^*}^{\dag}).
\end{equation}

Again using concavity of $S$ we have

\begin{equation}\label{}  S(\Phi^c(\rho_1\otimes \rho_2))\geq \sum_{\a} p_{\a} S((U_{\a} \rho_1
{U_{\a}}^{\dag}) \otimes (U_{\a}^* \rho_2 {{U_{\a}}^*}^{\dag}))
=\sum_{\a} p_{\a} (S(\rho_1)+S(\rho_2))=0,
\end{equation}
with equality only if all the states $U_{\a}\rho_1 U_{\a}^{\dagger}$
and ${U^*}_{\a}\rho_2 U_{\a}^T$ are independent of the index $\a$.
\\ This is possible only if we can find a pure state $|\psi\ra$
which transforms to the same state $|\phi\ra$ under the action of
all the operators $U_{\a}$, i.e. if
\begin{equation}\label{}
    \exists \ \ |\psi\ra : \hskip 3mm  U_{\a}|\psi\ra=e^{ic'_{\a}}|\phi\ra\h \forall \a,
\end{equation}
where $c'_{\a}$ is an arbitrary phase.  This last condition is
however exactly equivalent to the existence of a state $|\psi\ra$
which is invariant modulo a phase under all the operators
$U_{\a}^{-1}U_{\a_0}$ for an arbitrarily chosen $\alpha_{0}$, i.e.
 \begin{equation}\label{}
    U_{\a}^{-1}U_{\a_0}|\psi\ra=e^{ic_{\a}}|\psi\ra.
\end{equation}

Therefore if $\bigcap_{\a} I_{\a}=\O$, no separable state can
achieve zero entropy at the output of the channel and a critical
value of $\mu$ certainly exists.\\

Note that this theorem by itself does not imply that such a
transition is sharp, however the examples of \cite{M2,M3,C} and
those presented here suggest such a conclusion. As an example, we
are certain that in two dimensions, a channel with Kraus operators
$I,X, Z$ will show a transition while a
channel with Kraus operators $X, Z$ will not, a result which we have seen in our numerical
searches of optimal states.  \\

\section{Correlated channels with entanglement-enhanced
capacity}\label{capacity}

If we impose some extra conditions on the Kraus operators and follow
the arguments in \cite{M3}, we can show that minimum output entropy
is equivalent to maximal mutual information in the channel. Thus for
these kinds of channels the conditions stated in our
theorem also guarantee an enhancement of mutual information.\\

Following \cite{M3} we note that the first term of  mutual
information (\ref{mutual}), is maximized if
$\sum_ip_i\mathcal{E}(\rho_i)=\frac{1}{d^2}I$. Therefore an upper
bound is found for the mutual information in the form
\begin{equation}\label{bound}
I_2\leq 2\log_2 d-S(\mathcal{E}(\rho^*)).
\end{equation}

Let us suppose that the  Kraus operators in equation (\ref{Kraus})
have the following two properties: (a) commute with each other
modulo a phase: $U_aU_{a'}=U_{a'}U_a e^{i\phi_{\a,\a'}}\ \ $ , and
(b) form an irreducible
representation of a group .\\
Using property (a) and defining an equiprobable input ensemble
$\{p_{\a,\a'}={\rm const}, \rho_{\a,\a'}\}$ in which
$$\rho_{\a \a'}:=(U_{\a}\otimes U^*_{\a'})\rho^*(U_{\a}\otimes
U^*_{\a'})^{\dag}$$ it is straightforward to show that for
$\mathcal{E}$ defined in equation (\ref{Erho}) we have:
\begin{equation}
\mathcal{E}(\rho_{\a \a'})=(U_{\a}\otimes
U^*_{\a'})\mathcal{E}(\rho^*)(U_{\a}\otimes U^*_{\a'})^{\dag}.
\end{equation}
This mean that the channel $\mathcal{E}$ in covariant with
respect to the Kraus operators $U_{\a}$. Since entropy is
invariant under unitary operations, we conclude that
\begin{equation}\label{sterm}
S(\mathcal{E}(\rho_{\a \a'}))=S(\mathcal{E}(\rho^*))
\end{equation}
Therefore if $\rho^*$ minimizes the output entropy all the
$\rho_{\a \a'}$ do that, too.\\

As a consequence of the second property of the Kraus operators, we
find that the state $\sum_{\a,\a'}p_{\a,\a'}
\mathcal{E}(\rho_{\a,\a'})$ with all $p_{\a,\a'}$'s equal, commutes
with all the operators $U_{\a}\otimes U_{\a}^{\dagger}$ and hence by
Schur's first lemma
 it is a multiple of identity,
\begin{equation}\label{fterm}
\mathcal{E}(\sum_{\a \a'} p_{\a,\a'}\rho_{\a \a'})=\frac{1}{d^2}I
\end{equation}
From (\ref{mutual}), (\ref{sterm}) and (\ref{fterm}) we see that the
upper bound of equation (\ref{bound}) is attainable if we use
$\rho_{\a \a'}$ with the same probability as the input states. This
means that to maximize the mutual information, we only need to find
a state $\rho^*$ which minimizes the output entropy. \\

The examples which we present in the following sections are of this
type and hence the transition observed in their behavior as measured
in the minimum output entropy is in fact a transition in their
mutual information.

\section{A qubit channel with correlated noise}\label{qubit}

In \cite{M2} and \cite{M3} two examples of qubit channels which
have such a critical correlation have been studied.  We introduce
a third example. Let us take the error operators to be
\begin{equation}\label{IXZ}
U_1=I \hskip 1cm U_2=\sigma_x \hskip 1cm U_3=\sigma_z \hskip 1mm ,
\end{equation}
where $I$ is the identity matrix and $\sigma_x$ and $\sigma_z$ are
the Pauli matrices. These errors happen with probability $p$, $q$
 and $r$ respectively, with $p+q+r=1$. This is a channel with only
bit-flip and phase-flip operators.

\begin{figure}
  \centering
  \epsfig{file=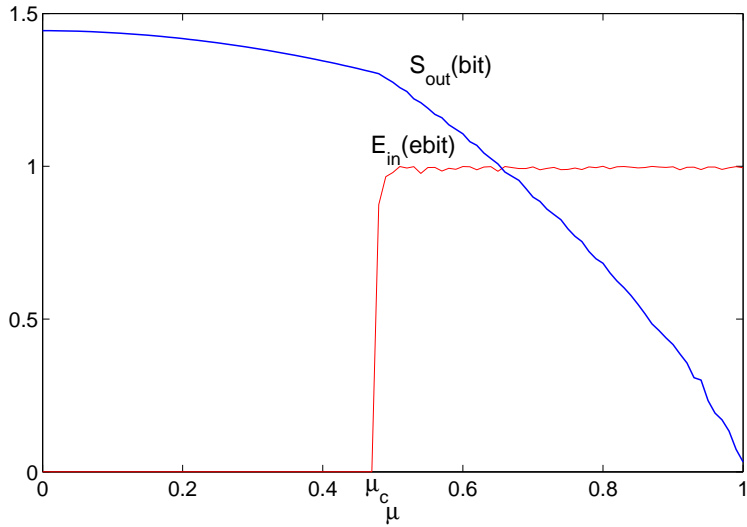,width=12cm}
  \caption{(Color online) Minimum output entropy and entanglement of the
           related input state as a function of $\mu$, for $p=0.3$, $q=0.2$ and $r=0.5$.
           }
  \label{min-ent-1}
\end{figure}
It is easily verified that this channel satisfies the criteria
mentioned in the previous section. We expect this channel to show
a transition as we increase the correlation parameter $\mu$. To
see this we take a general pure state of two qubits
\begin{equation}
|\psi\ra=a_0|00\ra+a_1|01\ra+a_2|10\ra+a_3|11\ra,
\end{equation}
and calculate the output state. The eigenvalues of the output
density matrix can not be determined analytically and have to be
evaluated numerically for fixed correlation parameter $\mu$ and
error parameters $p$ and $q$ for all states. For each set of these
parameters we find the optimal state, i.e. the input state which
yields the output state with minimum entropy. The result is shown in
figure (\ref{min-ent-1}) where we have fixed error parameters to
$p=0.3$ and $ q=0.2$.\\

It is seen that there is a sharp transition for optimal input state
at $\mu_c=0.47$ from separable states to maximally entangled states.

\section{Pauli channels in arbitrary dimensions}\label{arbitrary}

In this section we analyze correlated Pauli channels in arbitrary
dimensions and study in more detail a 3 dimensional channel, which
shows such a  critical transition.\\ In general for a generalized
Pauli channel carrying $d$ dimensional states (with basis states
$|0\ra,\cdots |d-1\ra$), the error operators are the generalized
Pauli operators $U_{mn}$ defined as
\begin{equation}\label{}
    U_{mn}|k\ra:=\xi^{kn}|k+m\ra,
\end{equation}
where $\xi:=e^{\frac{2\pi i}{d}}$. These operators have well-known
properties, including
\begin{eqnarray}\label{Uprop}
  U_{mn}^{\dagger} &=& \xi^{mn}U_{-m,-n},\cr
  U_{kl}U_{mn} &=& \xi^{lm-kn} U_{mn}U_{kl} ,\cr
  tr(U_{mn}) &=& d\delta_{m,0}\delta_{n,0},
\end{eqnarray}

and satisfy the conditions in section (\ref{capacity}). The
effect of a Pauli channel on a single qudit is defined as
\begin{equation}\label{dpauli}
    \Phi(\rho)=\sum_{m,n=0}^{d-1} p_{m,n}U_{mn}\rho
    U^{\dagger}_{mn},
\end{equation}
with $\sum_{m,n}p_{m,n}=1$.

In order to simplify the calculations we can restrict ourselves to a
subclass of such channels which have a symmetry.

Let us assume that such a channel has a symmetry of the form

\begin{equation}\label{}
    \Phi(S_{\a}\rho S_{\a}^{\dagger})=\Phi(\rho),
\end{equation}
for $\a $ belonging to an index set representing the symmetry group
$G$. Then if $\rho^*$ is an optimal state,
$\tilde{\rho}:=\frac{1}{|G|}\sum_{\a} S_{\a}\rho^* S_{\a}^{\dagger}$
will also be an optimal state, where $|G|$ is the order of the
group. Moreover the state $\tilde{\rho}$ is invariant, that is
$S\tilde{\rho} S^{\dagger} = \tilde{\rho}\ \ \ \ \forall S\in G$ .
This invariance greatly facilitates our analytical or numerical
search for optimal states.
\\ In the Pauli channel let us consider a subclass for which
$p_{m,n}=p_m$. This is a generalization of the channel studied in
\cite{M3}. In this subclass we have the following symmetry
\begin{equation}\label{symm}
    \Phi(U_{0k}\rho U_{0k}^{\dagger})=\Phi(\rho)\ \h \forall k.
\end{equation}
This symmetry also exists when the channel (\ref{dpauli}) acts on
two qudits in the presence of correlated noise. In that case it
takes the form
\begin{equation}\label{}
   \mathcal{E}(U_{0k}\otimes U^*_{0k})\rho(U_{0k}\otimes U^*_{0k})^{\dagger})=\mathcal{E}(\rho).
\end{equation}
Since $U_{0k}=(U_{01})^k$ this symmetry is generated by only one
single element namely $U_{01}$ and we can search for the optimal
input state among those which have the following symmetry.

\begin{equation}\label{}
   (U_{01}\otimes U^*_{01}) \rho^*(U_{01}\otimes
   U^*_{01})^{\dagger}=\rho^*.
\end{equation}

A simple calculation in the basis in which $U_{01}$ is diagonal
shows that the state $\tilde{\rho}$ is nothing but a convex
combination of pure states of the following form

\begin{equation}\label{pure}
    |\psi_{k}\ra:=\sum_{j=0}^{d-1}a_j|j,j-k\ra,\h k=0,1,\cdots d-1.
\end{equation}

Following the reasoning of \cite{M3} we can take the optimal state
to be a pure state which we take to be of the form  $|\psi_0\ra =
\sum_{j=0}^{d-1}a_j|j,j\ra$, without loss of generality. To find
the output state we calculate the correlated and uncorrelated
parts of the channel separately. For the correlated part we find
\begin{eqnarray}\label{Phic}
    \Phi^c(|\psi_0\ra)&=&\sum_{m,n,i,j}p_m a_i a_j^* (U_{mn}\otimes
    U_{mn}^*)|i,i\ra\la j,j|(U_{mn}\otimes
    U_{mn}^*)^{\dagger}\cr
    &=& \sum_{m,i,j} dp_m a_i a_j^* |i+m,i+m\ra\la j+m,j+m|.
\end{eqnarray}
For the uncorrelated part we have
\begin{equation}\label{}
    (\Phi\otimes \Phi^*)(|\psi_0\ra)=\sum_{i,j}a_ia_j^*
    \Phi(|i\ra\la j|)\otimes \Phi^*(|i\ra\la j|).
\end{equation}
Since
\begin{eqnarray}\label{}
    \Phi(|i\ra\la j|) &=& \sum_{m}p_m U_{mn}|i\ra \la
    j|U_{mn}^{\dagger} = \sum_{m,n}p_m \xi^{(i-j)n}|i+m\ra \la
    j+m|\cr &=&
    d\delta_{ij}(\sum_{m}p_m |i+m\ra \la i+m|).
\end{eqnarray}

we find
\begin{equation}\label{product}
    (\Phi\otimes \Phi^*)(|\psi_0\ra)= d^2 \sum_{i,m,n} p_mp_n|a_i|^2
    |i+m,i+n\ra \la i+m,i+n|.
\end{equation}

As in (\ref{Erho}) the complete output of the channel will be

\begin{equation}\label{E}
    \mathcal{E}(|\psi_0\ra)=(1-\mu)(\Phi\otimes
    \Phi^*)(|\psi_0\ra)+\mu\Phi^c(|\psi_0\ra).
\end{equation}

One can now determine the entropy of this output state and minimize
it with respect to the coefficients $a_i$ to determine the optimal
input state and its entanglement. This part of the problem must
usually be carried out numerically. We have done this task for 3
level states (qutrits), where the matrix of error parameters take
the form
\begin{equation}\label{}
    p_{m,n}\equiv\left(\begin{array}{ccc} p& p & p \\ q& q& q \\ r & r &
    r\end{array}\right).
\end{equation}
\begin{figure}
  \centering
  \epsfig{file=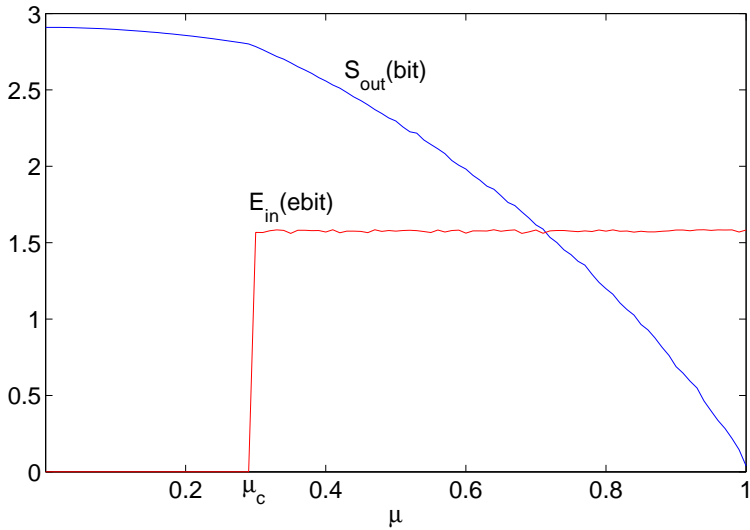,width=12cm}
  \caption{(Color Online) The minimum output entropy and the entanglement of
           the optimal state as a function of $\mu$ for a 3 dimensional symmetric Pauli
           channel. The critical value of $\mu$ is $\mu_c\approx 0.29$ for
           $p=0.0800$ ,$q=0.1800$ and $r=0.0733$.}\label{IST-Symmetry}
\end{figure}
For fixed error parameters, and for variable values of the
correlation parameters $\mu$, we have searched numerically among
all the states which minimize the output entropy. Figure
(\ref{IST-Symmetry}) shows the entropy of the output state when
the optimal state is fed into the channel. For each $\mu$ the
entanglement of the optimal input state is also plotted. It is
clearly seen that there is a sharp transition at $\mu_c\approx
0.29$. Below $\mu_c$ the optimal state is a separable state and
above $\mu_c$ it is a maximally entangled state. This plot is
typical, changing the error parameters only changes the value of
critical correlation $\mu_c$. Note that in calculating the
entanglement of the input state we have used logarithms to base 2
so that a maximally entangled state
has an entanglement of $\log_2{3}$.\\

The interesting features are that first, the transition is sharp and
not smooth and second no matter what the error parameters are, it is
the maximally entangled states and not some other states with lower
values of entanglement which minimize the output entropy and hence
maximize the mutual information. Therefore the transition is governed by a struggle of the
two extremes of entanglement. \\

Before concluding the paper it is instructive to compare the
fidelity of the output and input states and also the linearized
entropies of the output state for two extreme cases, namely when the
input state is completely separable and when it is a maximally
entangled state. This will provide us with a very  simple way to
obtain an estimate of the critical value of correlation.  \\

For the maximally entangled state $\rho^{ME}:=|\psi\ra\la \psi|$
where $|\psi\ra=\frac{1}{\sqrt{d}}\sum_i|i,i\ra$ , we put
$a_i=\frac{1}{\sqrt{d}}$ in (\ref{Phic}) and (\ref{product}) and
find
\begin{equation}\label{}
    \mathcal{E}(\rho^{ME}) = (1-\mu) d \sum_{m,n} C_{m,n} |m,n\ra \la
    m,n|+\mu \rho^{ME}
\end{equation}

where
\begin{equation}\label{Cmn}
 C_{mn}=\sum_{i}p_{m+i}p_{n+i}.
\end{equation}

 For a separable state
$\rho^s:=|0,0\ra\la 0,0|$ we find

\begin{equation}\label{}
    \mathcal{E}(\rho^{s}) = (1-\mu) \chi\otimes \chi +\mu d\sum_{m} p_{m} |m,m\ra \la
    m,m|,
\end{equation}
where $\chi=d\sum_m p_m |m\ra\la m|$. (Other separable states like
$|k,k\ra$ give different output states but the same output entropy).

\begin{figure}
  \centering
  \epsfig{file=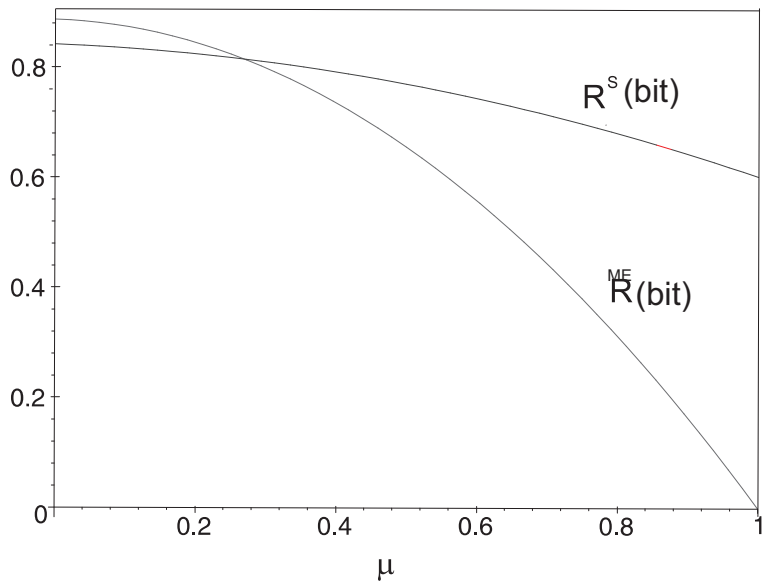,width=10cm}
  \caption{The linearized output entropies for the maximally entangled states ($R^{ME}$) and separable states
   $(R^s)$ as a function of $\mu$.}
  \label{purity}
\end{figure}

\begin{figure}
  \centering
  \epsfig{file=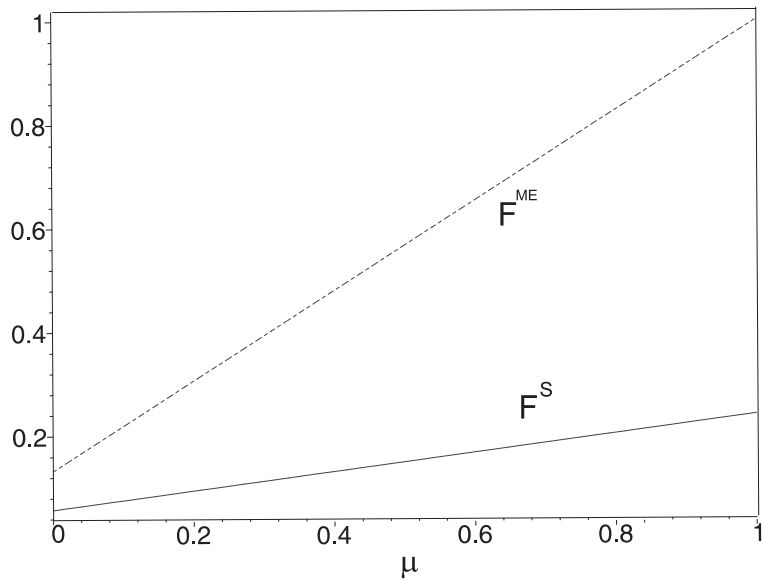,width=10cm}
  \caption{The fidelity of the output and input states for the maximally entangled states ($F^{ME}$) and separable states
   $(F^s)$, as a function of $\mu$. The parameter $\mu$ and the fidelities are dimensionless.}\label{fidelity}
\end{figure}

For the maximally entangled state the fidelity will be

\begin{equation}\label{}
    F^{ME}:=\la \psi|\mathcal{E}(\rho^{ME})|\psi\ra = \mu +
    (1-\mu)\sum_{n}C_{nn}=\mu+(1-\mu)d\sum_{n}p_n^2.
\end{equation}
and the linearized entropy
$R(\rho^{ME})=1-tr(\mathcal{E}^2(\rho^{ME}))$ will be
\begin{equation}
R^{ME}=1-[(1-\mu)^2d^2\sum_{m,n}C_{mn}^2+\mu^2+2\mu(1-\mu)dC_{00}]
\end{equation}
 while for separable states the corresponding quantities will be
\begin{equation}\label{}
    F^s=\la 0,0|\mathcal{E}(\rho^s)|0,0\ra = (1-\mu) d^2 p_0^2 + \mu d p_0.
\end{equation}

and
\begin{equation}
R^s=1-[(1-\mu)^2d^4C_{00}^2+\mu^2d^2C_{00}+2\mu(1-\mu)d^3\sum_{m}p_m^3].
\end{equation}
For the 3 dimensional channel that we have studied with the
parameters $p_{0}=0.08,\ \ p_1=0.18$ and $p_2=0.073$ we have plotted
in figure (\ref{purity}) and (\ref{fidelity}) the linearized
entropies of these two output states and their fidelity with their
input states.\\

We see that the maximally entangled states have a higher fidelity
than separable states at the output for all values of $\mu$, however
their output linearized entropy becomes less than that of the
separable states at almost the same critical value of
$\mu_c\approx0.28$ which we found by considerations of minimum
von-Neumann entropy.\\

Therefore it may be possible to analyze more complicated channels,
those without symmetry and in arbitrary dimensions, in a much
simpler way, i.e. by searching for optimal states either numerically
or analytically, according to the minimality of their linearized and
not von-Neumann entropy.

\section{discussion}
We have provided conditions under which a channel with correlated
noise shows a sharp transition in the form of its optimal states, as
the level of correlation passes a critical value. The interesting
point is that the transition occurs from completely separable states
to maximally entangled states and not states with some intermediate
value of entanglement, depending on the values of error parameters.
This phenomenon is reminiscent of phase transitions. In the same way
that in phase transitions there is a struggle between order and
disorder, i.e. between energy and entropy, here there is a struggle
between maximal entanglement and complete separability. One is
tempted to link this to a symmetry breaking phenomenon. In
ferromagnetic phase transitions we know that the free energy changes
its shape when we lower the temperature below the critical
temperature, and a unique minimum (with zero magnetization)
bifurcates to a a manifold of minima (with non-zero magnetization).
Is there a similar function here defined on the space of states or
their entanglement which undergo a similar change when we increase
the level of correlation? We think that this question deserves much
further investigation.

\section{Acknowledgement}
We would like to thank the members of the Quantum information group
of Sharif University for very valuable comments. Our special thanks
go to M. Dehghan Niri for providing generous help with some needed
softwares.

\end{document}